# A DFT investigation of methanolysis and hydrolysis of triacetin

Taweetham Limpanuparb[a], Kraiwan Punyain[b], Yuthana Tantirungrotechai[c,d,*]

[a]Department of Chemistry, Faculty of Science, Mahidol University, Bangkok 10400, Thailand
[b]Department of Chemistry, Faculty of Science, Naresuan University, Phitsanulok 65000, Thailand
[c]National Nanotechnology Center (NANOTEC), National Science and Technology Development Agency, Pathumthani 12120, Thailand
[d]Theoretical Chemistry Laboratory, Faculty of Science, Mahidol University, Salaya, Nakhon Pathom 73170, Thailand

* Corresponding author at: Address: National Nanotechnology Center (NANOTEC), National Science and Technology Development Agency, Pathumthani 12120, Thailand.

## Abstract

The thermodynamic and kinetic aspects of the methanolysis and hydrolysis reactions of glycerol triacetate or triacetin, a model triacylglycerol compound, were investigated by using Density Functional Theory (DFT) at the B3LYP/6-31++G(d,p) level of calculation. Twelve elementary steps of triacetin methanolysis were studied under acid-catalyzed and base-catalyzed conditions. The mechanism of acid-catalyzed methanolysis reaction which has not been reported yet for any esters was proposed. The effects of substitution, methanolysis/hydrolysis position, solvent and face of nucleophilic attack on the free energy of reaction and activation energy were examined. The prediction confirmed the facile position at the middle position of glycerol observed by NMR techniques. The calculated activation energy and the trends of those factors agree with existing experimental observations in biodiesel production.

## 1. Introduction

Triacylglyceride (TG) is a major constituent of naturally available oil and fat. The methanolysis, or transesterification, of triacylglyceride yields fatty acid methyl ester (FAME) and free glycerol as products. Alkaline methanolysis of triacylglycerol is a conventional process for biodiesel production from fresh vegetable oil owing to a very high conversion rate. This process makes use of a basic homogeneous catalyst such as KOH or NaOH [1]. The conversion rate is satisfactory provided that water and free fatty acid (FFA) contents in raw material are low [1]. This prerequisite prevents the hydrolysis reaction, the most important competing reaction, which yields soap as a product.

Methanolysis of triacylglycerol can be catalyzed by acid but with slower rate of conversion. Under acidic condition, the soap production is avoided and the esterification of FFAs which are found with high percentage in waste vegetable oil can take place simultaneously. However, this generally slow conversion rate of acid-catalyzed transesterification requires a rather high temperature condition [2].

We believe that the basic understanding on the methanolysis and its competing hydrolysis reaction of triacylglycerol compounds at the molecular level is essential for the development of biodiesel production. With the use of glycerol triacetate or triacetin as triacylglycerol model compound, we investigate the reaction mechanism of the methanolysis and its competing hydrolysis reaction under

acid- or base-catalyzed conditions. The triacetin molecule was considered in this work because it is the smallest compound in triacylglycerol family that contains all features of triacylglycerols. It undergoes three successive reactions until it becomes glycerol and releases three fatty acids or fatty acid methyl esters depending on the involved reaction. In this work, four types of reaction are considered: base-catalyzed hydrolysis (BH), base-catalyzed methanolysis (BM), acid-catalyzed hydrolysis (AH) and acid-catalyzed methanolysis (AM). There exists proposed reaction mechanism for BH, BM and AH reactions of some small compounds in the literature [3], [4], [5], [6], [7], [8], [9] and [10]. They are all bimolecular, base- or acid-catalyzed, acyl-oxygen cleavage reactions proceeded via the formation and dissolution of a tetrahedral intermediate. Schematically shown for the based-catalyzed hydrolysis in Fig. 1, there are five important stationary structures along the reaction coordinate: reactant complex (RC), first transition state (TS1), tetrahedral intermediate (TI), second transition state (TS2) and product complex (PC) [7] and [8].

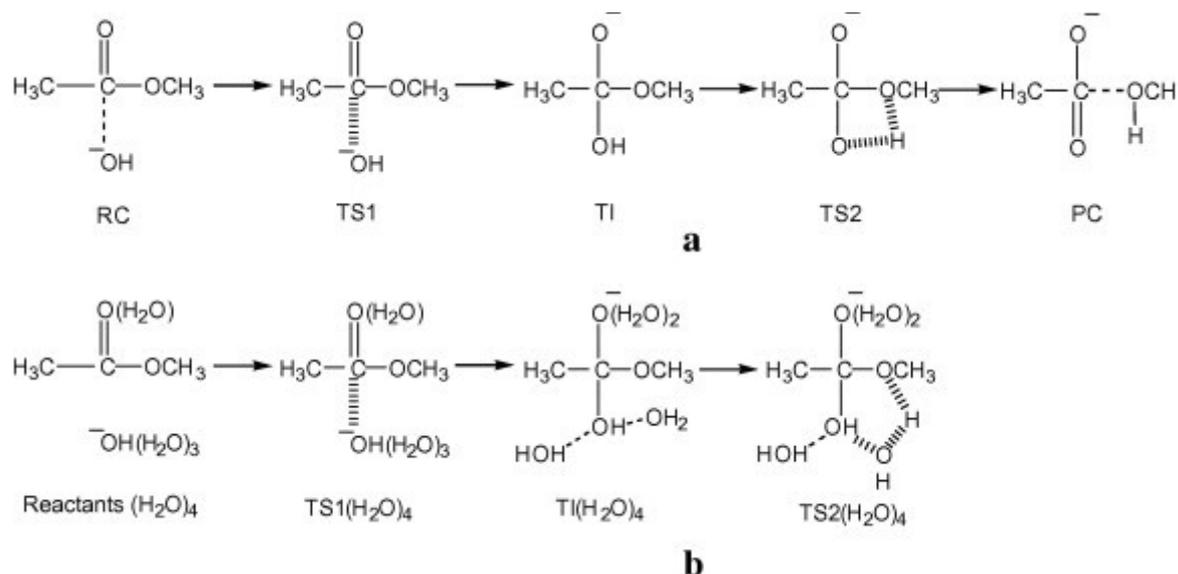

Fig. 1. Schematic reaction mechanism of the based-catalyzed hydrolysis of methyl acetate (a) without and (b) with explicit water molecules proposed by Zhan et al. [7] and [8]. The hydrogen bond shown by dash line helps facilitate this reaction.

The reaction mechanism of the based-catalyzed hydrolysis of esters has long been studied by several groups [3], [4], [5], [6], [7], [8] and [9]. Zhan et al. reported the hydrolysis of methyl acetate in gas phase and solution phase with the solvents being represented explicitly or by the solvation model [7] and [8]. In gas phase, five stationary structures along the reaction coordinate are identified with the dissolution of tetrahedral intermediate (TI) being the rate-determining step. This rate-determining step agrees with the experimental observation [11]. On the other hand, with the combination of four explicit solvent molecules and implicit solvent, only four stationary structures are reported. It was found that the RC structure in solvent phase is less stable than the separated reactants. It was then excluded from the calculation and the energy barrier was taken directly from the difference in energy of TS1 and separated reactants [6] and [8]. Zhan and Landry further reported the alkaline ester hydrolysis of cocaine and revealed that the face of nucleophilic attack can play a role on the reaction kinetics [6]. The energy barrier difference for the attack at the Si and Re faces of cocaine molecule is ∼1 kcal/mol.

At least two reports exist on the mechanism of base-catalyzed methanolysis of methyl acetate based on the previous hydrolysis work of Zhan and coworkers [12] and [13]. The dissolution of tetrahedral intermediate does not facilitate by the proton transfer as in the hydrolysis counterpart. Theoretical studies of alkaline methanolysis of vegetable oil in the literature are rather limited. Om Tapanes et al. conducted an experiment and theoretical studies of biodiesel formation from *Jatropa curcas* oil [14]. With the intention to clarify the existence of one or two tetrahedral intermediates along the reaction pathways of base-catalyzed methanolysis, the authors investigated the reaction pathways of monoglyceride by using the semi-empirical AM1 model. Only one tetrahedral intermediate was observed in their calculation. The dissolution of this tetrahedral intermediate was found to be the rate-determining step. The authors attribute the difference between the kinetics of the methanolysis and the ethanolysis to the alkoxide formation step. They also assume some similarity between each successive stage of conversion from triglyceride to glycerol. Asakuma et al. consider the difference between each stage of conversion in details [15]. Three different successive pathways for the base-catalyzed transesterification of various vegetable oil were proposed and investigated by using the HF/STO-3G method. Several alkoxides were considered in their work. The activation energy was found to be decreasing with the alkoxide size. Almost no variation of the activation energy with the carbon chain length of fatty acid was observed. The authors conclude that middle ester bond in the triglyceride is transesterified before the ester bond at the end position.

Recently Hori et al. reported the mechanism of gas-phase acid-catalyzed hydrolysis of methyl acetate in gas phase and solution phase modeled by explicit and implicit solvents [10]. The authors reported that the reaction proceeds only if when an explicit solvent molecule is included (see Fig. 2). In gas phase, the dissolution of TI is the rate-determining step. This was confirmed experimentally by several kinetic studies [16], [17], [18], [19] and [20]. They noticed that explicit solvent molecules enhance the nucleophilicity of the attacking water which is a weaker nucleophile than hydroxide ion and facilitate the leaving of $-OCH_3$ group. Without additional water molecule, no stable TI structure was obtained [10]. According to this gas-phase mechanism, the barrier height of the second step is greater than that of the first step, implying that the leaving of $-OCH_3$ group with water-assisted proton transfer is the rate-determining step.

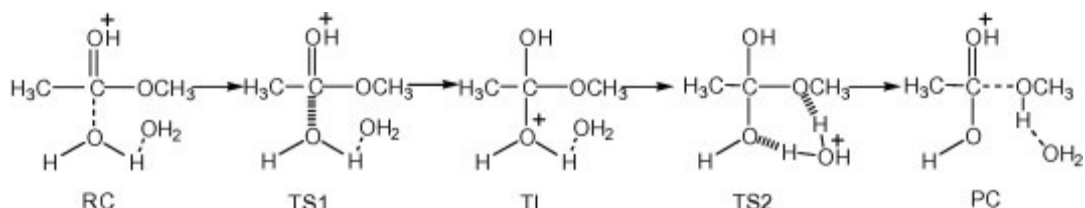

Fig. 2. Schematic reaction mechanism of the acid-catalyzed hydrolysis of methyl acetate as proposed by Hori et al. [10].

To the best of our knowledge, the reaction mechanism of the acid-catalyzed methanolysis has not yet been elucidated in the literature. Fox and coworkers reported an unsuccessful attempt to calculate acid-catalyzed methanolysis mechanism of methyl acetate [13]. They concluded that the methanol molecule could not get close to the carbonyl carbon enough to form a TI structure. This might be due to the neglect of explicit solvent molecule as in the case of acid-catalyzed hydrolysis reaction investigated by Hori group [10]. Therefore, we explored a plausible mechanism of acid-catalyzed methanolysis reaction in this work.

## 2. Computational details

Glycerol triacetate or triacetin, an ester of glycerol and three acetic acids, was chosen as a triacylglycerol model compound. As observed by Asakuma et al., the carbon chain length of fatty acid plays minor role in the kinetics [15]. Therefore, the smallest acid side chain renders triacetin a practical model compound for computational study of biodiesel formation. It is utilized as a model compound in recent kinetic investigations of acid- and base-catalyzed methanolysis which aim to develop better catalysts for biodiesel synthesis [21] and [22]. These studies provide us with activation energies that can be used as a benchmark for our investigation.

Because triacetin can adopt several conformers, we conducted a preliminary search on the conformational energy landscape of triacetin. The two-dimensional conformational space based on two dihedral angles around the C–C bonds of the glycerol backbone was explored. Fig. 3 displays the mentioned dihedral angles and the numbering scheme used on the glycerol backbone. Nine minima and 18 first-order transition states of triacetin were identified on a two-dimensional conformational space. The most stable conformer is the one with two backbone OCCO torsional angles being *trans* and *gauche*. Our finding contradicted with reported MM2 result which predicted that both dihedral angles should be gauche plus and gauche minus for the global minimum [23]. We adopt the DFT global minimum structure of triacetin as an initial geometry for our study.

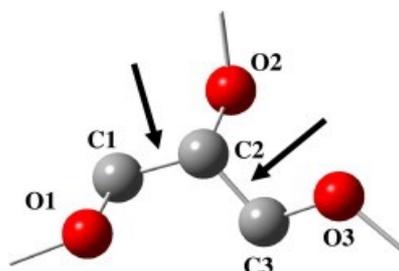

Fig. 3. The glycerol backbone of the most stable triacetin structure (H points outward at C2).

To elucidate the methanolysis/hydrolysis reactions of triacetin, we partitioned the whole reaction into three successive steps as originally outlined by Yamasaki [24]. Fig. 4 illustrates 12 possible elementary steps to methanolyze or hydrolyze three ester linkages of triacylglyceride. Each step is essentially the removal of single acyl group by attacking nucleophile of which the reaction mechanism proceeds with tetrahedral intermediate as stated earlier. The form of nucleophile depends on the catalyst used. For acid-catalyzed reaction, the nucleophile is the neutral $CH_3OH$ or $H_2O$ molecule for methanolysis and hydrolysis reactions, respectively. During the reaction, the protonation occurs on the carbonyl oxygen of the ester substrate. For base-catalyzed reaction, we consider the $CH_3O^-$ or $OH^-$ ions as nucleophile for methanolysis and hydrolysis reactions, respectively. The nucleophile of this category also acts at the same time as basic catalyst. In both cases, the reaction proceeds through the nucleophilic substitution, then followed by either protonated ester or negatively charged nucleophile regeneration.

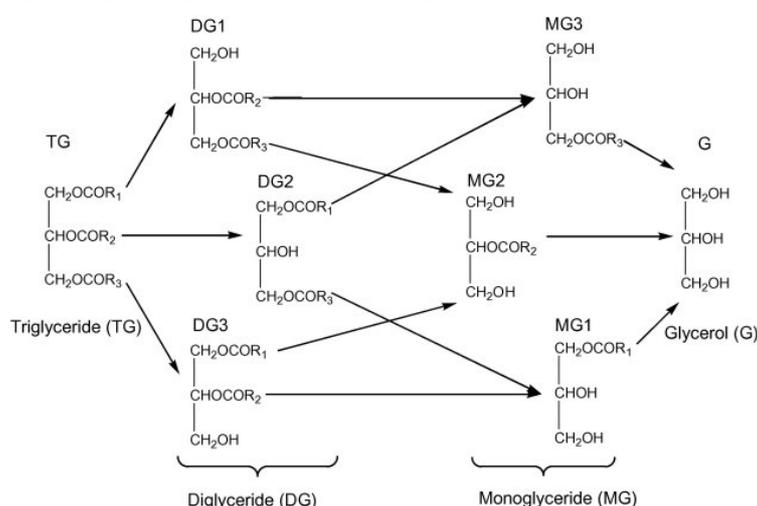

Fig. 4. All possible successive transformation steps for the methanolysis and hydrolysis of triacylglyceride to glycerol and three fatty acids or fatty acid methyl esters.

The geometries of all compounds in the scheme as well as that of RC, TS1, TI, TS2 and PC for all elementary steps were optimized by using the Density Functional Theory (DFT) at B3LYP/6-31++G(d,p) level. All stationary structures were verified by computing their hessian. The CPCM continuum solvation model was used to incorporate the solvent effect. In some case, explicit solvent molecules were included in the calculations. The reaction free energy and the activation energy for all 12 elementary steps were evaluated and compared with available experimental kinetics data of triacetin and vegetable oil [22], [25], [26] and [27]. All calculations in this work were performed by using the Gaussian 03 program [28].

# 3. Results and discussion

Table 1 summarizes average reaction free energies for the methanolysis and hydrolysis reactions of triacetin in the acid- and base-catalyzed conditions. The reaction free energies were averaged from those of all 12 possible elementary steps (see Fig. 4 and Table 2). The hydrolysis of triacetin is slightly exergonic in gas phase (−0.46 kcal/mol) but becomes moderately exergonic in solution phase (∼−4 kcal/mol). On the other hand, the reaction free energies for triacetin methanolysis are approximately the same for gas and solution phase (∼−2 kcal/mol). From thermodynamics point of view, methanolysis is, therefore more favorable than hydrolysis in gas phase, but becomes less favorable than hydrolysis in solution phase. This discrepancy between gas and solution phase suggests that the solvents play an important role and should be included in the calculation to better reproduce experimental data. Although the reaction free energies in acid- and base-catalyzed conditions are the same, the reaction free energies of the nucleophilic substitution in these two conditions are different. The corresponding energy in basic condition is greater in magnitude than those in acid condition. This is probably due to the anionic nature of involving species. The choice of either water or methanol solvents simulated by CPCM model hardly affects the reaction free energy.

Table 1. Average reaction free energies for hydrolysis and methanolysis of triacetin (in kcal/mol). The reaction free energies were averaged over 12 possible elementary steps.

| <ΔG>/(kcal/mol) | Gas phase | CPCM energy at gas phase geometries | |
|---|---|---|---|
| | | Water | Methanol |
| *Acid-catalyzed reactions* | | | |
| (1) Protonation | −32.53 | −5.72 | −6.04 |
| Hydrolysis reaction | **−0.46** | **−3.71** | **−3.52** |
| (2) Nucleophilic substitution | 9.53 | −5.86 | −5.23 |
| (3) Protonated ester regeneration | −9.99 | 2.15 | 1.71 |
| Methanolysis reaction | **−2.15** | **−2.80** | **−2.69** |
| (2) Nucleophilic substitution | −0.63 | −4.19 | −3.70 |
| (3) Protonated ester regeneration | −1.53 | 1.38 | 1.02 |
| *Base-catalyzed reactions* | | | |
| Hydrolysis reaction | **−0.46** | **−3.71** | **−3.52** |
| (1) Nucleophilic substitution | −45.19 | −28.47 | −28.59 |
| (2) Nucleophile regeneration | 44.73 | 24.76 | 25.07 |
| Methanolysis | **−2.15** | **−2.80** | **−2.69** |
| (1) Nucleophilic substitution | −21.02 | −8.93 | −9.23 |
| (2) Nucleophile regeneration | 18.87 | 6.13 | 6.54 |

Bold values signifies that these numbers are sum or average of the other.

Table 2. Reaction free energies for 12 elementary steps of triacetin methanolysis (in kcal/mol) and their averages. The reaction free energies were averaged according to (i) the number of acyl substituents on glycerol backbone and (ii) the position on the glycerol backbone that methanolysis takes place.

|  | Gas phase | CPCM water | | CPCM methanol | |
| --- | --- | --- | --- | --- | --- |
|  |  | Solvent phase structure | Gas phase structure | Solvent phase structure | Gas phase structure |
| MeOAc | 0.00 | 0.00 | 0.00 | 0.00 | 0.00 |
| TG→DG1 | −1.56 | −3.09 | −3.02 | −2.28 | −2.81 |
| TG→DG2 | −3.41 | −4.33 | −4.72 | −4.14 | −3.75 |
| TG→DG3 | −2.45 | −1.23 | −2.44 | −2.85 | −1.91 |
| DG1→MG2 | −1.69 | −1.82 | −1.88 | −1.40 | −1.61 |
| DG1→MG3 | −2.98 | −2.90 | −3.69 | −4.74 | −2.86 |
| DG2→MG1 | −1.36 | −0.19 | −1.76 | −0.87 | −2.37 |
| DG2→MG3 | −1.14 | −1.66 | −1.99 | −2.88 | −1.92 |
| DG3→MG1 | −2.32 | −3.29 | −4.03 | −2.16 | −4.20 |
| DG3→MG2 | −0.80 | −3.68 | −2.46 | −0.83 | −2.51 |
| MG1→G | −2.40 | −2.46 | −2.06 | −2.13 | −2.06 |
| MG2→G | −3.91 | −2.07 | −3.63 | −3.46 | −3.75 |
| MG3→G | −2.62 | −0.99 | −1.82 | −0.12 | −2.51 |
| *No. of acyl substituent* | | | | | |
| TG→DG | −2.41 | −2.81 | −3.40 | −3.09 | −2.82 |
| DG→MG | −1.65 | −2.18 | −2.65 | −2.15 | −2.58 |
| MG→G | −2.91 | −1.76 | −2.52 | −1.90 | −2.77 |
| *Position of methanolysis* | | | | | |
| 1 | −1.41 | −2.65 | −2.40 | −2.03 | −2.33 |
| 2 | −3.08 | −3.07 | −4.03 | −3.63 | −3.64 |
| 3 | −1.96 | −0.98 | −1.99 | −1.31 | −2.10 |
| Average | **−2.15** | **−2.23** | **−2.80** | **−2.32** | **−2.69** |

Bold values signifies that these numbers are sum or average of the other.

Table 2 also reports the effect of optimized geometries on the reaction free energy. By using the gas-phase and solution-phase optimized geometries, the reaction free energies including the CPCM solvation are generally very consistent. We therefore adopted the gas phase optimized geometries to reduce computational costs for the rest of this work.

The variation of the reaction free energies of 12 elementary methanolysis steps in Table 2 is believed to be due to three factors: (1) the number and position of acetyl substituents on glycerol backbone, (2) the nucleophilic attacking position on glycerol backbone that methanolysis or hydrolysis takes place, and (3) the solvent effect. These thermodynamic quantities computed in gas and solution phase in Table 2 can also be considered as the relative stability of triacetin derivatives with respect to glycerol and three methyl acetate molecules (see also Fig. S1 in Supplementary information). The reaction free energy of an elementary step depends on the stability of reactants and products which in turn depends on their structures, i.e. position that reaction takes place, and the number and position of acetyl substituents.

Judging from the reaction free energies, the gas phase reactions follow these trends: MG → G > TG → DG > DG → MG and position 2 > position 3 > position 1. In other word, the MG → G is the most exergonic reaction compared to TG → DG and DG → MG processes and is the most thermodynamically favorable process. Regarding the substitution position on the glycerol backbone, the nucleophile prefers to attack at the position 2 or the middle position. On the other hand, in solution phase the reaction follows these trends: TG → DG > DG → MG ∼ MG → G and position 2 > position 1 > position 3. The trend in the attacking position on glycerol backbone in the gas phase reflects the steric hindrance in triacylglyceride molecule, i.e. the position 2 or the middle position is the most steric. The greater the steric relief is, the more exergonic the reaction becomes. For the side positions, the position 3 is slightly more steric than the position 1 because the acyl substituent at positions 3 and 1 of glycerol backbone is *gauche* and *trans* to the middle group, respectively. This trend at the side position is reversed in the solution phase. In solution, we believe that the dipole moment determines this trend. In polar solvents, the acyl group at the position 1, which is *trans* to the middle position, results in partial cancellation of the dipole moment of the reactant, hence making it less stable than those with acyl group at the position 3.

The hydrolysis and methanolysis reactions under both acid- and base-catalyzed conditions can be divided into major steps depending on the reaction condition (see Table 1). Under acid condition, the ester is protonated first before undergoing a nucleophilic substitution. Finally, a product is deprotonated by transferring a proton to another ester reactant and, therefore, regenerating a reactive ester to undergo nucleophilic substitution in the next cycle. Under basic condition, the negatively charged nucleophile directly attacks the ester. Then the product abstracts a proton from reactants to regenerate negatively charged nucleophile again. As the proton transfer process is known to be very fast process, our focus is on the nucleophilic substitution which is believed to be rate-determining step. Although the total reactions should have the same reaction free energy independent of the catalysts involved, the acid-catalyzed and base-catalyzed reactions differ greatly in the nucleophilic substitution step. Considering the reaction energy of this step, the hydrolysis and methanolysis reactions are more favorable in the base-catalyzed condition than in the acid-catalyzed condition. This observation is in line with the experimental observation that rate of methanolysis of vegetable oil in alkaline medium is about 4000 times faster than that in acidic medium [2]. However, to compare with these experimental observations, the reaction energy barrier in acid- and base-catalyzed conditions must be obtained.

In the following subsections, detailed reaction mechanisms of the nucleophilic substitution step in acid- and base-catalyzed conditions, including the corresponding energy barriers, will be discussed.

### 3.1. Acid-catalyzed methanolysis (AM)

Two possible mechanisms for acidic methanolysis are considered. Based on a report that tetrahedral intermediate of the reaction could not be obtained [13], a one-step mechanism with a transition structure was calculated as illustrated in Fig. 5. The barrier height in both gas and solution phases is over 30 kcal/mol. This is unusually high and almost three times greater than the experimental value of ∼11 kcal/mol for triacetin [22]. We thus proposed two-step mechanism comprising of RC, TS1, TI, TS2 and PC stationary structures. Initially, we tried to optimize the TI structure consisting of only the ester and attacking nucleophile. The attempt was unsuccessful because the nucleophile could not get close to the ester enough to form the TI structure. Our finding agrees with what reported previously in the literature [10] and [13]. Therefore, we adopted the approach of Hori et al. which adds an explicit solvent molecule to enhance nucleophilicity of nucleophile and to assist proton transfer [10]. Using this technique, we were able to calculate the reaction mechanism of acid-catalyzed methanolysis for methyl acetate as well as for all 12 elementary steps of triacetin (see Table 3). By setting the energy of RC structure as the reference, the average free energy of five stationary structures over 12 elementary steps is depicted in Fig. 6 along with the stationary structures of MG1 → G step as an illustrative example. The dissolution of TI structure is the rate-determining step of this mechanism.

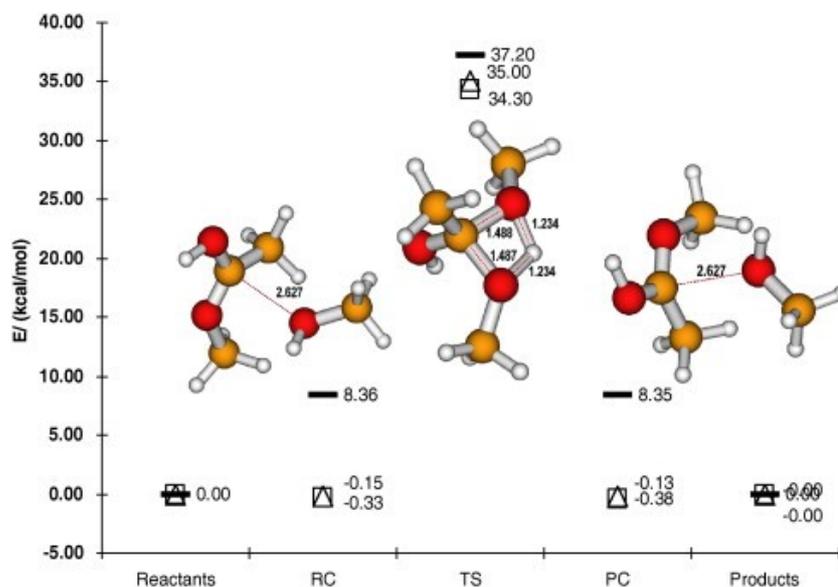

Fig. 5. Free energy reaction profile of one-step acid-catalyzed methanolysis mechanism in gas phase (−), water (□) and methanol (△).

Table 3. The free energy reaction profile of triacetin acid-catalyzed methanolysis in gas phase (in kcal/mol). The reaction proceeds through five stationary structures as outlined in Fig. 6. The face of nucleophilic attack is indicated in parenthesis.

|  | RC → TS1 | →TI | →TS2 | →PC | Activation energy ($E_a$) |
|---|---|---|---|---|---|
| MeOAc | 11.24 | −2.12 | 5.82 | −16.64 | 14.94 |
| TG → DG1 (Re) | 11.63 | −3.09 | 7.47 | −43.07 | 16.01 |
| TG → DG2 (Re) | 7.92 | −0.77 | 3.10 | −36.59 | 10.25 |
| TG → DG3 (Re) | 12.49 | −2.91 | 7.98 | −43.95 | 17.56 |
| DG1 → MG2 (Re) | 12.93 | −2.11 | 6.28 | −44.31 | 17.11 |
| DG1 → MG3 (Re) | 6.66 | −1.71 | 6.06 | −44.00 | 11.00 |
| DG2 → MG1 (Re) | 12.19 | −0.97 | 7.48 | −43.58 | 18.70 |
| DG2 → MG3 (Re) | 11.29 | −2.02 | 6.54 | −37.11 | 15.81 |
| DG3 → MG1 (Re) | 6.92 | 0.20 | 5.70 | −38.86 | 12.82 |
| DG3 → MG2 (Re) | 10.40 | 0.01 | 2.33 | −34.13 | 12.73 |
| MG1 → G (Re) | 11.89 | −3.21 | 6.16 | −37.36 | 14.84 |
| MG2 → G (Re) | 8.38 | −0.89 | 6.35 | −39.08 | 13.84 |
| MG3 → G (Re) | 12.58 | −2.17 | 7.11 | −39.84 | 17.52 |
| MG1 → G (Si) | 11.55 | −2.44 | 7.62 | −41.21 | 16.73 |
| MG2 → G (Si) | 8.62 | −4.90 | 9.40 | −33.51 | 13.13 |
| MG3 → G (Si) | 14.30 | −3.76 | 4.25 | −36.36 | 14.79 |

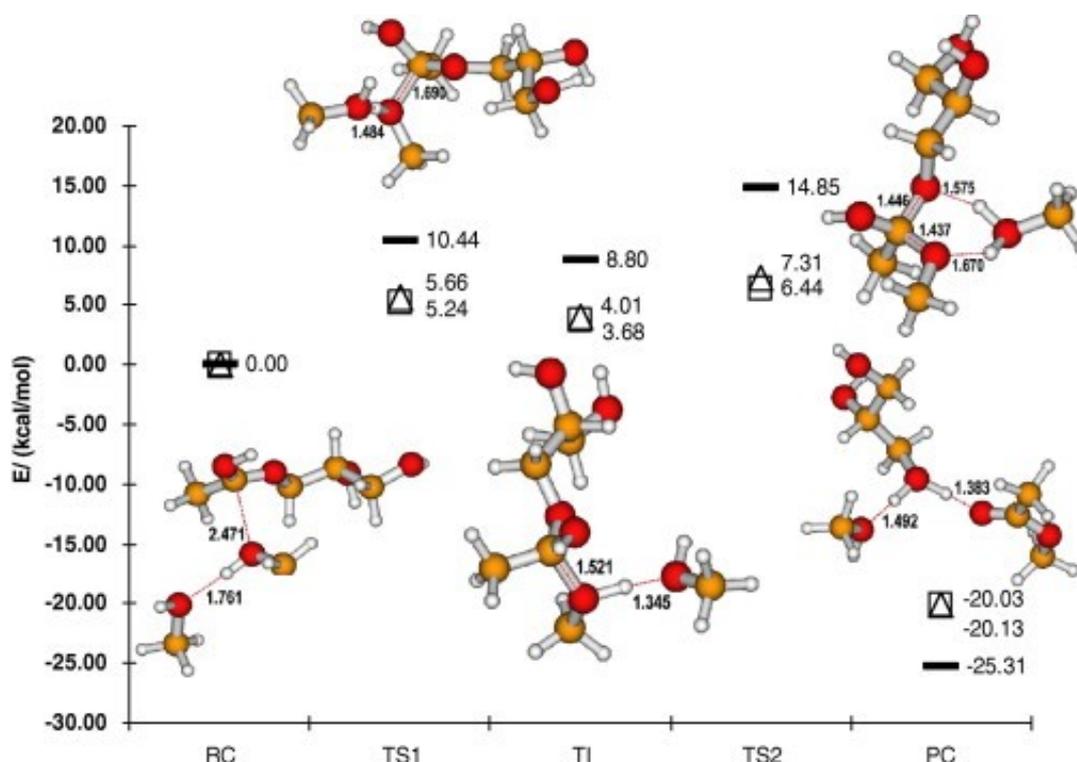

Fig. 6. Free energy reaction profile of acid-catalyzed methanolysis averaged over 12 elementary steps in gas phase (−), water (□) and methanol (△) environments. The geometries of MG1 undergoing acid-catalyzed methanolysis (Re face attack of methanol) are shown as a representative of the reaction.

Table 6 reports the activation energy of the triacetin methanolysis averaged over 12 possible elementary steps for different number of acetyl groups on glycerol backbone and different nucleophilic attacking position. The mechanism provides us with the activation energy of ∼15 kcal/mol in gas phase and ∼7 kcal/mol in solution phase. According to the table, the DFT underestimated the barrier energies in solution phase; this is due to the shortcoming of DFT on proton transfer process [29] and [30]. We therefore carried out Hartree–Fock energy calculation on DFT geometries for comparison. The experimental TG → DG activation energy of ∼11 kcal/mol was found to lie in between average TG → DG DFT and HF barrier heights of ∼8 and ∼14 kcal/mol.

Table 6. Average activation free energies for the acid-catalyzed and base-catalyzed methanolysis (in kcal/mol) in gas phase and solvent phase. The experimental values are given for comparison.

|  | Acid-catalyzed reaction | | | | Base-catalyzed reaction | | | Experimental value |
|---|---|---|---|---|---|---|---|---|
|  | Gas phase | CPCM solvent | | | Gas phase | CPCM solvent | |  |
|  |  | Water | MeOH | HF[a]:MeOH |  | Water | MeOH |  |
| *No. of acyl substituent* | | | | | | | | |
| TG → DG | 14.60 | 6.88 | 7.83 | 14.12 | 2.33 | 9.62 | 10.54 | 11.03[b], 11.71[c], 14.7[d], 7.57[e] |
| DG → MG | 14.69 | 6.08 | 6.86 | 13.16 | 1.48 | 10.26 | 10.86 | 18.44[c], 14.2[b], 9.94[e] |
| MG → G | 15.40 | 7.10 | 7.95 | 14.35 | 1.63 | 7.53 | 7.93 | 7.94[c], 6.4[d], 1.42[e] |

|  | Acid-catalyzed reaction | | | | Base-catalyzed reaction | | | Experimental value |
|---|---|---|---|---|---|---|---|---|
|  | Gas phase | CPCM solvent | | | Gas phase | CPCM solvent | | |
|  |  | Water | MeOH | HF[a]:MeOH |  | Water | MeOH | |
| *Position of methanolysis* | | | | | | | | |
| 1 | 14.85 | 5.73 | 6.65 | 13.91 | 2.99 | 6.89 | 7.40 | |
| 2 | 11.98 | 6.52 | 7.09 | 11.68 | 4.21 | 12.31 | 13.20 | |
| 3 | 17.72 | 7.36 | 8.39 | 15.49 | 2.30 | 9.05 | 9.55 | |
| Overall | **14.85** | **6.54** | **7.38** | **13.70** | **3.16** | **9.42** | **10.05** | |

[a]HF, Hartree–Fock.
[b]Triacetin/methanol/$H_2SO_4$ [22].
[c]Soybean oil/methanol/NaOH [26].
[d]Palm oil/methanol/KOH [25].
[e]Sunflower oil/methanol/KOH [27].

In gas phase, the preferential nucleophilic attacking position predicted from the reaction free energy agrees well with those predicted from the energy barrier. The ester bond at the middle position is the most reactive and also releases the most energy. In this case, the more thermodynamically favorable a reaction is, the faster it proceeds. However, in solution phase, the acidic methanolysis at position 1 has the least activation energy. This suggests that a reaction in solvent is likely to occur at a position which is less sterically hindered, so that a nucleophile can easily approach the carbonyl carbon. To our knowledge, there has not been any experiment under acidic condition that considers the nucleophilic attacking position factor yet. Our prediction, therefore, remains to be tested by experiment.

Another factor that could affect the activation energy of acid-catalyzed methanolysis is the face of nucleophilic attack. This factor has already been reported to be rather small in the case of base-catalyzed hydrolysis [6]. We considered the acid-catalyzed methanol attack on the Si and Re faces of MG1. As reported in Table 3, the difference in barrier height between the attack on Si and Re faces is about 3 kcal/mol in gas phase and 2 kcal/mol is solution phase.

## 3.2. Acid-catalyzed hydrolysis (AH)

By adopting initial geometry of stationary structures from AH mechanism reported for methyl acetate [10], the acidic hydrolysis mechanism of MG1 was summarized in Fig. 7. The rate-determining step is the dissolution of TI which is similar to the acid-catalyzed methanolysis mechanism. The MG1 activation energy for AH reaction of about 13 kcal/mol is comparable to the experimental activation energy of methyl acetate which is reported to be 16 kcal/mol [31].

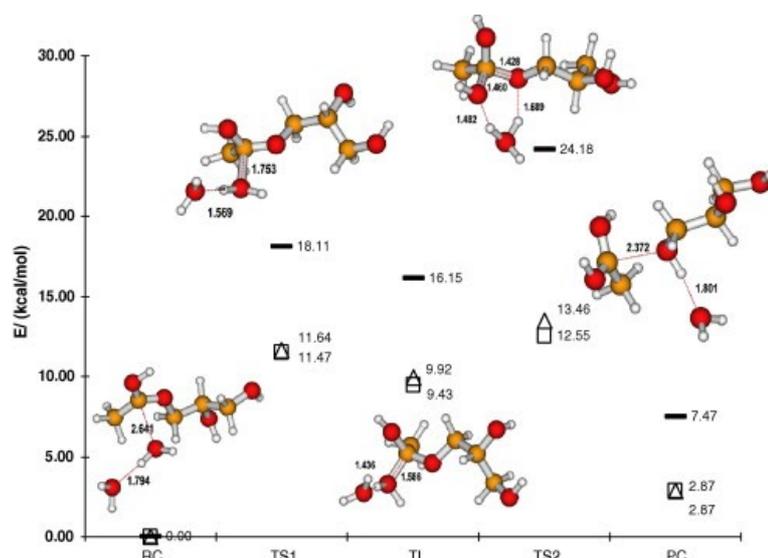

Fig. 7. Free energy reaction profile of MG1 acid-catalyzed hydrolysis (Re face attack of water) in gas phase (−), water (□) and methanol (△) environments.

As there are several factors that affect acid-catalyzed methanolysis and hydrolysis mechanism, the role of nucleophile, explicit and implicit solvent molecule on the activation energy were reported in Table 4. The activation energy of acid-catalyzed hydrolysis process is greater than that of methanolysis, implying that methanol is a better nucleophile than water. Methanol is also a better facilitating solvent compared to water as it results in a lower activation energy. The acid-catalyzed methanolysis of triacetin hence proceeds considerably faster than its corresponding hydrolysis reaction judging from these two observations. An implicit water model decreased the activation energy by about 1 kcal compared to those calculating with the methanol model. This reduction is however small compared to 3–7 kcal/mol difference originated from the choice of nucleophile and explicit solvent. Therefore the explicit solvent treatment, even with the minimal amount of solvent molecules, is a prerequisite to a correct description of this reaction.

Table 4. The activation energies for acid-catalyzed reaction of methyl acetate (MeOAc) and MG1 calculated with different nucleophile, facilitating solvent and implicit CPCM solvent. The energies were reported in terms of Gibbs free energy (in kcal/mol).

| Ester | Gas phase | Implicit Solvent | |
| Nucleophile/explicit solvent | | Water | MeOH |
|---|---|---|---|
| MeOAc | | | |
| MeOH/MeOH | 14.94 | 6.54 | 7.41 |
| MeOH/H$_2$O | 20.81 | 9.49 | 10.50 |
| H$_2$O/MeOH | 18.19 | 10.83 | 11.20 |
| H$_2$O/H$_2$O | 24.61 | 12.45 | 13.24 |
| MG1 → G (Re) | | | |
| MeOH/MeOH | 14.84 | 6.51 | 7.61 |
| MeOH/H$_2$O | 21.01 | 9.61 | 10.92 |
| H$_2$O/MeOH | 17.31 | 9.83 | 10.37 |
| H$_2$O/H$_2$O | 24.18 | 12.55 | 13.46 |

As the dissolution of TI structure coupled with the water-assisted proton transfer is the rate-determining step for both acid-catalyzed methanolysis (AM) and hydrolysis reactions (AH) of triacetin, the proton tunneling effect that was neglected so far might significantly reduce the calculated activation energy of this step. A rough estimation of the proton tunneling correction obtained from the Wigner expression at 298.15 K is less than 1 kcal/mol [6], [32] and [33]. This number is in agreement with an earlier study on base-catalyzed hydrolysis (BH); it also confirms that the second step of AM and AH is still the rate-determining step as the proton tunneling correction is too small to reverse the relative degree of barrier heights of these two steps [8].

### 3.3. Base-catalyzed methanolysis (BM)

The stationary structures and their corresponding energy along the reaction coordinate for 12 successive steps of the base-catalyzed methanolysis of triacetin were obtained with the help from mechanisms reported by Fox et al. and Chen and Brauman [12] and [13]. Table 5 summarizes the free energy reaction profile of triacetin base-catalyzed methanolysis in gas phase while Table 6 reports the activation energy in both gas and solution phases. In gas phase, the relative energy of five stationary points along the reaction coordinates is similar to those in the acid-catalyzed reaction. However the reactant complex (RC) structure becomes less stable than reactants in solution phase. We excluded the reactant complex from the energy barrier calculation of the first step and considered the energy difference between TS1 and separated reactants (SR) structures instead (see Fig. 8).

Table 5. The free energy reaction profile of triacetin base-catalyzed methanolysis in gas phase (in kcal/mol). The reaction proceeds through five stationary structures as outlined in Fig. 8. The face of nucleophilic attack is indicated in parenthesis.

|  | RC → TS1 | →TI | →TS2 | →PC | Activation energy ($E_a$) |
|---|---|---|---|---|---|
| MeOAc | 2.11 | −4.50 | 4.50 | −2.09 | 2.11 |
| TG → DG1 (Re) | 2.00 | −9.82 | −1.62 | −6.10 | 2.00 |
| TG → DG2 (Si) | 4.27 | −14.71 | −2.62 | −4.01 | 4.27 |
| TG → DG3 (Si) | 1.93 | −7.40 | −1.58 | −0.57 | 1.93 |
| DG1 → MG2 (Si) | 1.95 | −15.11 | 10.74 | −7.56 | 1.95 |
| DG1 → MG3 (Si) | 5.96 | −11.17 | −4.13 | −3.19 | 5.96 |
| DG2 → MG1 (Re) | 3.39 | −7.25 | 1.91 | −7.71 | 3.39 |
| DG2 → MG3 (Re) | 5.28 | −7.05 | 2.56 | −2.49 | 5.28 |
| DG3 → MG1 (Si) | 6.47 | −18.84 | 6.07 | −15.15 | 6.47 |
| DG3 → MG2 (Re) | 1.84 | −12.01 | 3.25 | −4.24 | 1.84 |
| MG1 → G (Re) | 2.83 | −2.52 | 0.68 | −3.91 | 2.83 |
| MG2 → G (Si) | 0.14 | −10.14 | 3.02 | −13.02 | 0.14 |
| MG3 → G (Re) | 1.91 | −3.42 | 3.27 | −4.20 | 1.91 |

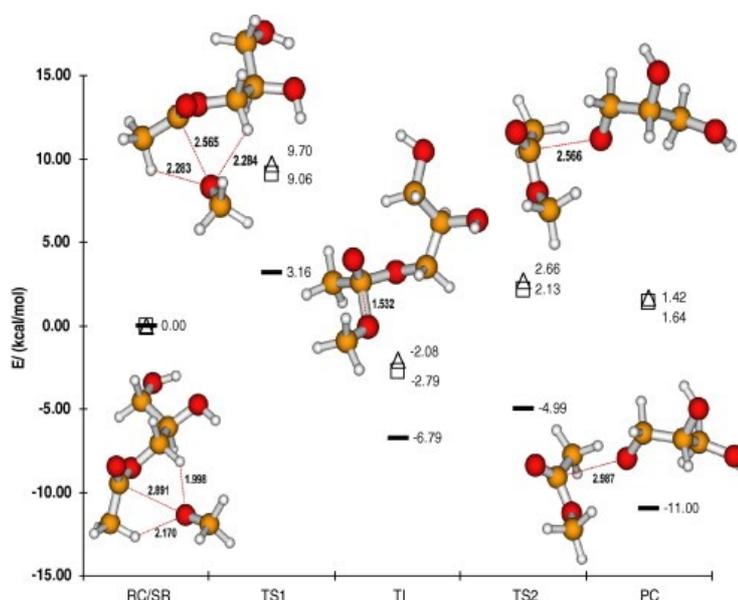

Fig. 8. Free energy reaction profile of base-catalyzed methanolysis averaged over 12 elementary steps in gas phase (−), water (□) and methanol (△) environments. The geometries of MG1 undergoing base-catalyzed methanolysis (Re face attack of methoxide ion) are shown as a representative of the reaction.

In contrast to the acid-catalyzed reaction, the formation of TI structure is the rate-determining step. According to the barrier height in Table 6, the calculations of this reaction were in qualitative agreement with the experimental observations (Refs. b and d in Table 6); the activation energy $E_a$ follows this trend: DG → MG > TG → DG > MG → G [26] and [27]. Although the experimental activation energy of 1–18 kcal/mol were obtained from triacylglyceride with long-chain fatty acids [25], [26] and [27], our calculations on triacetin yield comparable values of 8–11 kcal/mol. These values fall into the range of activation energy observed in other triacylglycerides. We believe that triacetin is justified to be used as model compound representing the other triacylglycerides including those used as feedstock for biodiesel production.

### 3.4. Base-catalyzed hydrolysis (BH)

The optimized geometries and energies of MG1 species undergoing the base-catalyzed hydrolysis were calculated as illustrative case based on the proposed mechanism of methyl acetate by Zhan et al. [6], [7] and [8]. The energy profiles in gas and solution phases were depicted in Fig. 8. The separated reactants energies were used to calculate the barrier height for the same reason as the base-catalyzed methanolysis reaction.

There are two additional solvated stationary structures, TI (water) and TS2 (water), for water-assisted dissolution of tetrahedral intermediate. An explicit water molecule plays an important role in reducing the energy barrier in this step. This finding confirms what previously reports by Zhan et al. [6], [7] and [8]. The reaction proceeds via this alternative pathway with the formation of TI structure being the rate-determining step. According to Fig. 8, the energy barrier of 10.19 kcal/mol in water agrees well with the experimental activation energy of 10.45 kcal/mol for the alkaline hydrolysis of methyl acetate in pure water by Fairclough and Hinshelwood [11]. The authors reported also that the activation energy of the same reaction in solution containing 71.3% ethanol in water (w/w) is 15.0 kcal/mol. This indicates that the calculated solvent effect agrees qualitatively with the experimental observation that the barrier height increases as one increase the alcohol composition in solution.

# 4. Conclusions

We conducted series of DFT investigation for acid- and base-catalyzed hydrolysis and methanolysis reactions involving in biodiesel synthesis by using the triacetin as a model compound. Two acid-catalyzed methanolysis mechanisms were proposed and one-step mechanism with tetrahedral complex as the transition structure was ruled out based on the experimental observation. Then the mechanisms of 12 elementary steps of the acid- and base-catalyzed triacetin methanolysis were studied in detail. The calculated activation energies averaged over 12 elementary steps for different number of acetyl groups on triacetin and for the nucleophilic attacking position on the glycerol backbone are comparable to the available experimental values. The effects of the number of acetyl groups on triacetin and the solvent on the base-catalyzed methanolysis agree with the experiments. The preferential attacking position of methanol at the middle position of the glycerol backbone agrees well with the NMR experiment recently conducted by Jin et al. [34]. The thermodynamic trend in gas phase generally follows the kinetic trend while in solution phase only the kinetic trend is observed experimentally.

The gas phase activation energies of these four types of reactions (see Table 6 and Fig. 6, Fig. 7, Fig. 8 and Fig. 9) support general observation found in biodiesel production, i.e. the reaction proceed faster in basic condition, and the hydrolysis is a very important competing reaction under base-catalyzed condition but is unlikely to occur under acid-catalyzed condition. In solution phase, the activation energy of acid-catalyzed reaction decreases, while that of base-catalyzed reaction increases compared to the gas-phase condition. As the barrier height of acid-catalyzed methanolysis is likely to be underestimated by the present DFT functional, the activation energy of acid- and base-catalyzed reactions in solution phase could not be directly compared. The Hartree–Fock method was employed and found to overestimate the barrier height. The experimental activation energy was therefore found to lie in between the DFT and HF activation energies. The reaction mechanisms reported here cover all 12 elementary steps for converting triacetin successively to free glycerol and three methyl acetates (or three acetic acids) and could be further modified to include a body of heterogeneous catalysts into account.

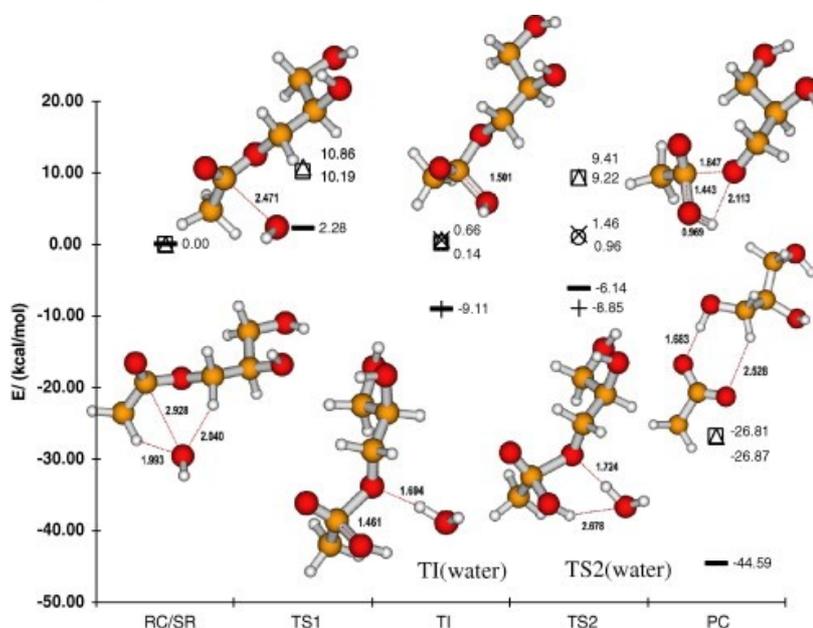

Fig. 9. Free energy reaction profile of MG1 base-catalyzed hydrolysis (Re face attack of hydroxide ion) in gas phase (−, +), water (□, ○) and methanol (△, ×) environments. The energies of water-assisted proton transfer (+, ○, ×) are shown along with explicit water associated TI (water) and TS2 (water) geometries.

# Acknowledgements

T.L. gratefully acknowledges the Junior Science Talent Project (JSTP Grant No. 06-50-4R) and the Development and Promotion of Science and Technology Talent Project (DPST Grant No. 08/2547) for scholarship and research funding. Y.T. acknowledges financial support from the National Nanotechnology center (NANOSIM) and Thailand Research Fund (RSA5180010). We thank the HPC service, NECTEC for computational resources.